# Cover Schemes, Frame-Valued Sets and Their Potential Uses in Spacetime Physics

John L. Bell

The concept of *Grothendieck (pre)topology* or *covering* issued from the efforts of algebraic geometers to study "sheaf-like" objects defined on categories more general than the lattice of open sets of a topological space (see, e.g. [4]). A *Grothendieck pretopology* on a category $\mathscr{C}$ with pullbacks is defined by specifying, for each object $U$ of $\mathscr{C}$, a set $P(U)$ of arrows to $U$ called covering families satisfying appropriate category theoretic versions of the corresponding conditions for a family $\mathscr{A}$ of sets to cover a set $U$, namely: (i) $\{U\}$ covers $U$, (ii) if $\mathscr{A}$ covers $U$ and $V \subseteq U$, then $\mathscr{A}|V = \{A \cap V : A \in \mathscr{A}\}$ covers $V$, and (iii) if $\mathscr{A}$ covers $U$ and, for each $A \in \mathscr{A}$, $\mathscr{B}_A$ covers $A$, then $\bigcup_{A \in \mathscr{A}} \mathscr{B}_A$ covers $U$. In the present paper the covering concept—here called a *cover scheme*—is presented and developed in the simple case when the underlying category is a preordered set. The relationship between cover schemes, frames (complete Heyting algebras), Kripke models, and frame-valued set theory is discussed. Finally cover schemes and frame-valued set theory are applied in the context of Markopoulou's [5] account of discrete spacetime as sets "evolving" over a causal set.

### I. COVER SCHEMES ON PREORDERED SETS

Let $(P, \leq)$ be a fixed but arbitrary preordered set: we shall use letters $p, q, r, s, t$ to denote elements of $p$. We write $p \cong q$ for $(p \leq q$ & $q \leq p)$. A *meet* for a subset $S$ of $P$ is an element $p$ of $P$ such that $\forall q[\forall s \in S(q \leq s) \leftrightarrow q \leq p]$: if $p$ and $p'$ are both meets for $S$, then $p \cong p'$. If the empty subset $\emptyset$ has a meet, any such meet $m$ is necessarily a *largest* or *top* element of $P$, that is, satisfies $p \leq m$ for all $p$. We use the symbol 1 to denote a top element of $P$. A meet of a finite subset $\{p_1,...,p_n\}$ of $P$ will be denoted by $p_1 \wedge ... \wedge p_n$. $P$ is a *lower semilattice* if each nonempty finite subset of $P$ has a meet. A subset $S$ of $P$ is said to be a *sharpening of,* or to *sharpen,* a subset $T$ of $P$, written $S \prec T$, if $\forall s \in S \exists t \in T(s \leq t)$. A *sieve* in $P$ is a subset $S$ such that $p \in S$ and $q \leq p$ implies $q \in S$. Each subset $S$ of $P$ generates a sieve $\overline{S}$ given by $\overline{S} = \{p : \exists s \in S(p \leq s)\}$.

A *cover scheme* on $P$ is a map $\mathbf{C}$ assigning to each $p \in P$ a family $\mathbf{C}(p)$ of subsets of $p\!\downarrow\, = \{q: q \leq p\}$, called (**C**-)*covers of p,* such that, if $q \leq p$, any cover of $p$ can be sharpened to a cover of $q$, i.e.,

(*) $\qquad S \in \mathbf{C}(p) \,\&\, q \leq p \to \exists T \in \mathbf{C}(q)[\forall t \in T \exists s \in S(t \leq s)]$.

If $P$ is a lower semilattice, a *coverage* (see [3]) on $P$ is a map $\mathbf{C}$ as above, satisfying, in place of (*), the condition

$$S \in \mathbf{C}(p) \,\&\, q \leq p \to S \wedge q = \{s \wedge q : s \in S\} \in \mathbf{C}(p).$$

A cover scheme $\mathbf{C}$ is said to be *normal* if every member of every $\mathbf{C}(p)$ is a sieve and whenever $S \in \mathbf{C}(p)$ and $T$ is a sieve such that $S \subseteq T \subseteq p\!\downarrow$, we have $T \in \mathbf{C}(p)$. Any cover scheme $\mathbf{C}$ on $P$ induces a normal cover scheme $\overline{\mathbf{C}}$ (called its *normalization*) defined by

$$\overline{\mathbf{C}}(p) = \{X \subseteq p\!\downarrow : X \text{ is a sieve } \& \exists S \in \mathbf{C}(p).\ S \subseteq X\}.$$

Notice that a normal cover scheme on a lower semilattice is always a coverage. For if $\mathbf{C}$ is such, then for $S \in \mathbf{C}(p)$ and $q \leq p$, any sharpening of $S$ to a member of $\mathbf{C}(q)$ is easily seen to be included in $S \wedge q$, so that the latter is also in $\mathbf{C}(q)$.

Write $\mathscr{C}\!ov(P)$ for the set of all cover schemes on $P$. There is a natural partial ordering $\triangleleft$ on $\mathscr{C}\!ov(P)$ defined by

$$\mathbf{C} \triangleleft \mathbf{D} \leftrightarrow \forall p\ \mathbf{C}(p) \subseteq \mathbf{D}(p).$$

With this ordering $\mathscr{C}ov(P)$ is a complete lattice in which the join $\bigvee_{i \in I} \mathbf{C}_i$ of any family $\{\mathbf{C}_i : i \in I\}$ is given by

$$(\bigvee_{i \in I} \mathbf{C}_i)(p) = \bigcup_{i \in I} \mathbf{C}_i(p).$$

There is also a natural *composition* ★ defined on $\mathscr{C}ov(P)$. For $\mathbf{C}, \mathbf{D} \in \mathscr{C}ov(P)$, $\mathbf{D} \star \mathbf{C}$ is defined by decreeing that $(\mathbf{D} \star \mathbf{C})(p)$ is to consist of all subsets of $p{\downarrow}$ of the form $\bigcup_{s \in S} T_s$, where $S \in \mathbf{C}(p)$ and, for each $s \in S$, $T_s \in \mathbf{D}(s)$. That $\mathbf{D} \star \mathbf{C}$ is a cover scheme on $P$ may be verified (using the axiom of choice) as follows. Given $S \in \mathbf{C}(p)$, $\bigcup_{s \in S} T_s \in (\mathbf{D} \star \mathbf{C})(p)$ and $q \leq p$, there is $U \in \mathbf{C}(q)$ with $U \prec S$, so for each $u \in U$ there is $s(u) \in S$ for which $u \leq s(u)$. Then $T_{s(u)} \in \mathbf{D}(s(u))$ and we can choose $V_u \in \mathbf{D}(u)$ so that $V_u \prec T_{s(u)}$. Clearly $\bigcup_{u \in U} V_u \in (\mathbf{D} \star \mathbf{C})(q)$ and, since $V_u \prec T_{s(u)}$ for all $u \in U$, it follows immediately that $\bigcup_{u \in U} V_u \prec \bigcup_{s \in S} T_s$.

It is not hard to verify that ★ is associative and that with this operation $\mathscr{C}ov(P)$ is actually a *quantale*, that is, for any $\mathbf{D}, \{\mathbf{C}_i : i \in I\}$ in $\mathscr{C}ov(P)$,

$$\mathbf{D} \star \bigvee_{i \in I} \mathbf{C}_i = \bigvee_{i \in I}(\mathbf{D} \star \mathbf{C}_i) \quad (\bigvee_{i \in I} \mathbf{C}_i) \star \mathbf{D} = \bigvee_{i \in I}(\mathbf{C}_i \star \mathbf{D}).$$

Also the element $\mathbf{1} \in \mathscr{C}ov(P)$ with $\mathbf{1}(p) = \{p\}$ acts as a quantal unit, since it is readily verified that $\mathbf{1} \star \mathbf{C} = \mathbf{C} \star \mathbf{1} = \mathbf{C}$ for all $\mathbf{C} \in \mathscr{C}ov(P)$.

In this connection a *Grothendieck pretopology*—which we shall abbreviate simply to *pretopology*—on $P$ may be identified as a cover scheme $\mathbf{C}$ on $P$ satisfying $\mathbf{1} \triangleleft \mathbf{C}$ and $\mathbf{C} \star \mathbf{C} \triangleleft \mathbf{C}$, that is, $\{p\} \in \mathbf{C}(p)$ for all $p \in P$ and, if $S \in \mathbf{C}(p)$ and, for each $s \in S$, $T_s \in \mathbf{C}(s)$, then $\bigcup_{s \in S} T_s \in \mathbf{C}(p)$.

We observe that a normal pretopology **C** has the additional properties: (i) each **C**($p$) is a filter of sieves in $p\downarrow$, that is, satisfies $S, T \in \mathbf{C}(p) \leftrightarrow S \in \mathbf{C}(p)\ \&\ T \in \mathbf{C}(p)$; (ii) $S \in \mathbf{C}(p)\ \&\ q \leq p \rightarrow S \cap q\downarrow \in \mathbf{C}(q)$. For (ii), we observe that $S \cap q\downarrow$, including as it does any sharpening of $S$ to a member of **C**($q$), is itself a member of **C**($q$). As for (i), the "$\rightarrow$" direction is obvious; conversely, if $S, T \in \mathbf{C}(p)$, then $S \cap T = S \cap \bigcup_{t \in T}(t\downarrow) = \bigcup_{t \in T}(S \cap t\downarrow)$. But from (ii) we have $S \cap t\downarrow \in \mathbf{C}(t)$ for every $t \in T$, whence $\bigcup_{t \in T}(S \cap t\downarrow) \in \mathbf{C}(p)$, and so $S \cap T \in \mathbf{C}(p)$.

A normal pretopology is also called a *Grothendieck topology*. A normal cover scheme satisfying (i) and (ii) is called a *regular* cover scheme.

Each cover scheme **C** generates a pretopology, and a Grothendieck topology in the following way. First, define $\mathbf{C}^n$ for $n \in \omega$ recursively by $\mathbf{C}^0 = \mathbf{1}$ and $\mathbf{C}^{n+1} = \mathbf{C} \star \mathbf{C}^n$. Now put $\mathbf{G} = \bigvee_{n \in \omega} \mathbf{C}^n$. Then **G** is a pretopology, for obviously $\mathbf{1} \triangleleft \mathbf{G}$, and

$$\mathbf{G} \star \mathbf{G} = \bigvee_{n \in \omega} \mathbf{C}^n \star \bigvee_{m \in \omega} \mathbf{C}^m = \bigvee_{n \in \omega}\bigvee_{m \in \omega} \mathbf{C}^{m+n} = \bigvee_{n \in \omega} \mathbf{C}^n = \mathbf{G}.$$

Also $\mathbf{C} \triangleleft \mathbf{G}$, and **G** is evidently the $\triangleleft$-least such pretopology. **G** is called the *pretopology generated by* **M**. The normalization $\overline{\mathbf{G}}$ of **G** is then a Grothendieck topology called the *Grothendieck topology generated by* **C**.

Now let **M** be a map assigning to each $p \in P$ a subset **M**($p$) of subsets of $p\downarrow$. Since $\mathscr{C}\!ov(P)$ is a complete lattice, there is a $\triangleleft$-least cover scheme **C** such that $\mathbf{M}(p) \subseteq \mathbf{C}(p)$ for all $p$. **C** is called the cover scheme *generated by* **M**; the pretopology and Grothendieck topology generated in turn by **C** are said to be *generated by* **M**.

There are several naturally defined cover schemes on $P$ which also happen to be pretopologies. First, each sieve $A$ in $P$ determines two cover schemes $\mathbf{C}_A$ and $\mathbf{C}^A$ defined by

$$S \in \mathbf{C}_A(p) \leftrightarrow p \in A \cup S \qquad S \in \mathbf{C}^A(p) \leftrightarrow p\!\downarrow \cap A \subseteq S:$$

these are easily shown to be pretopologies. Notice that $\emptyset \in \mathbf{C}_A(p) \leftrightarrow p \in A$ and $\emptyset \in C^A(p) \leftrightarrow p\!\downarrow \cap A = \emptyset$.

Next, we have the *dense cover scheme* **Den** given by:

(*) $$S \in \mathbf{Den}(p) \leftrightarrow \forall q \leq p \exists s \in S \exists r \leq s (r \leq q):$$

it is a straightforward exercise to show that this is a pretopology. When $S$ is a sieve, the above condition (*) is easily seen to be equivalent to the familiar condition of *density below $p$*: that is, $\forall q \leq p \exists s \in S(s \leq q)$.

Note that the following are equivalent for any cover scheme **C**: (a) $\mathbf{C} \triangleleft \mathbf{Den}$, (b) $\emptyset \notin \mathbf{C}(p)$ for all $p$. For since $\emptyset \notin \mathbf{Den}(p)$, (a) clearly implies (b). Conversely, assume (b), and let $S \in \mathbf{C}(p)$. Then for each $q \leq p$ there is $T \in \mathbf{C}(q)$ for which $\forall t \in T \exists s \in S (t \leq s)$. Since (by (b)) $T \neq \emptyset$, we may choose $t_0 \in T$ and $s_0 \in S$ for which $t_0 \leq s_0$. Since $t_0 \leq q$, and $q \leq p$ was arbitrary, it follows that $S$ satisfies the condition (*) above for membership in $\mathbf{Den}(p)$. This gives (a).

Finally, we have the *Beth cover scheme* **Bet**. This is defined as follows. First we define a *road* from $p$ to be a maximal linearly preordered subset of $p\!\downarrow$: clearly any road from $p$ contains $p$. Let us call a a *rome* over $p$ any subset of $p\!\downarrow$ intersecting every road from $p$. Now the Beth coverage has $\mathbf{Bet}(p)$ = collection of all romes over $p$. Let us check first that **Bet** is a cover scheme. Suppose that $S$ is a rome over $p$ and $q \leq p$. We claim that

$$T = \{t \leq q : \exists s \in S(t \leq s)\}$$

is a rome over $q$. For let $Y$ be any road from $q$; then, by Zorn's lemma, $Y$ may be extended to a road $X$ from $p$. We note that since $X \cap q\!\downarrow$ is linearly preordered and includes $Y$, it must coincide with $Y$. Since $S$ is a rome over $p$, there must be an element $s \in S \cap X$. Since also $q \in Y \subseteq X$, we have $s \leq q$ or $q \leq s$. If $s \leq q$, then $s \in X \cap q\!\downarrow = Y$ and $s \in T$, so that $s \in Y \cap T$. If $q \leq s$, then $q \in T$; since $q \in Y$, it follows that $q \in Y \cap T$. So in either case $Y \cap T \neq \emptyset$; therefore $T$ is a rome over $q$. Since clearly also $T \prec S$, we have shown that **Bet** is a cover scheme.

To show that **Bet** is a pretopology, we observe first that, for any $p$, $\{p\}$ is a rome over $p$. Now suppose that we are given: a rome $S$ over $p$, for each $s \in S$, a rome $T_s$ over $s$, and a road $X$ from $p$. Then $s \in X \cap S$ for some $s$: we claim that $X \cap s{\downarrow}$ is a road from $s$. For suppose $t \leq s$ is comparable with every member of $X \cap s{\downarrow}$; now since $s \in X$, for each $x \in X$ either $s \leq x$ or $x \leq s$. In the first case $t \leq x$; in the second $t$ is comparable with $x$ by assumption. Hence $t$ is comparable with every member of $X$, and so $t \in X$. Accordingly $X \cap s{\downarrow}$ is, as claimed, a road from $s$; as such, it must meet the rome $T_s$, so $X$ meets $\bigcup_{s \in S} T_s$, and the latter is therefore a rome over $p$. So **Bet** is indeed a pretopology.

Since clearly $\emptyset \notin \mathbf{Bet}(p)$ for any $p$, it follows from what we have noted above that **Bet** ◁ **Den**, a fact that can also be easily verified directly.

Any preordered set $(P, \leq)$ generates a *free lower semilattice* $\tilde{P}$ which may be described as follows. The elements of $\tilde{P}$ are the finite subsets of $P$; the preordering on $\tilde{P}$ is the *refinement* relation $\sqsubseteq$, that is, for $F, G \in \tilde{P}$,

$$F \sqsubseteq G \leftrightarrow \forall q \in G \exists p \in F (p \leq q).$$

The meet operation $\wedge$ in $\tilde{P}$ is set-theoretic union; the canonical embedding of $P$ into $\tilde{P}$ is the map $p \mapsto \{p\}$. Notice also that $\emptyset$ is the unique top element of $\tilde{P}$.

Now, suppose we are given a cover scheme **C** on $P$. This induces a cover scheme $\tilde{\mathbf{C}}$ on $\tilde{P}$ defined in the following way. We start by setting $\tilde{\mathbf{C}}(\emptyset) = \{\{\emptyset\}\}$. Now fix a nonempty finite subset $F$ of $P$, take any nonempty subset $\{p_1, ..., p_n\}$ of $F$ and any $S_1 \in \mathbf{C}(p_1), ..., S_n \in \mathbf{C}(p_n)$. Define

$$S_1 \bullet ... \bullet S_n = \{\{s_1, ..., s_n\} \cup F : s_1 \in S_1, ..., s_n \in S_n\}.$$

We decree that $\tilde{\mathbf{C}}(F)$ is to consist of all sets of the form $S_1 \bullet ... \bullet S_n$, for $S_1 \in \mathbf{C}(p_1), ..., S_n \in \mathbf{C}(p_n)$, and all nonempty finite subsets $\{p_1, ..., p_n\}$ of $F$.

Let us check that $\tilde{\mathbf{C}}$ is a cover scheme on $\tilde{P}$. To begin with, the unique cover $\{\varnothing\}$ of $\varnothing$ is clearly sharpenable to any cover $S_1 \bullet ... \bullet S_n$ of any nonempty member of $\tilde{P}$. Now suppose that $S_1 \bullet ... \bullet S_n$ is a $\tilde{\mathbf{C}}$-cover of a nonempty member $F$ of $\tilde{P}$ and that $G = \{q_1, ..., q_m\} \sqsubseteq F$. Then for each $1 \leq i \leq n$ there is $q_i \in G$ for which $q_i \leq p_i$, hence $T_i \in \mathbf{C}(q_i)$ with $T_i \prec S_i$. Clearly $T_1 \bullet ... \bullet T_n \in \tilde{\mathbf{C}}(G)$. Also $T_1 \bullet ... \bullet T_n \prec S_1 \bullet ... \bullet S_n$. For, given $t_1 \in T_1, ..., t_n \in T_n$, then since $T_i \prec S_i$ for each $i$, there are $s_1 \in S_1, ..., s_n \in S_n$ for which $t_1 \leq s_1, ..., t_n \leq s_n$, whence $\{t_1, ..., t_n\} \cup G \sqsubseteq \{s_1, ..., s_n\} \cup F$. So $\tilde{\mathbf{C}}$ satisfies the conditions of a cover scheme.

The normalization $\overline{\tilde{\mathbf{C}}}$ of $\tilde{\mathbf{C}}$ is then a coverage on $\tilde{P}$ called the coverage on $\tilde{P}$ *induced* by **C.**

Now we associate a complete Heyting algebra, a *frame* (see, e.g. [3]) with each cover scheme on $P$. First, we define $\hat{P}$ to be the set of sieves in $P$ partially ordered by inclusion: $\hat{P}$ is then a frame—the *completion*[1] of $P$— in which joins and meets are just set-theoretic unions and intersections, and in which the operations $\Rightarrow$ and $\neg$ are given by

$$I \Rightarrow J = \{p : I \cap p{\downarrow} \subseteq J\} \qquad \neg I = \{p : I \cap p{\downarrow} = \varnothing\}.$$

Given a cover scheme **C** on $P$, a sieve $I$ in $P$ is said to be **C**-*closed* if

$$\exists S \in \mathbf{C}(p)(S \subseteq I) \to p \in I.$$

---

[1] Writing **Lat** for the category of complete lattices and join preserving homomorphisms, $\hat{P}$ is in fact the object in **Lat** freely generated by $P$.

We write $\widehat{\mathbf{C}}$ for the set of all **C**-closed sieves in $P$, partially ordered by inclusion.

**Lemma.** If $I \in \widehat{P}$, $J \in \widehat{\mathbf{C}}$, then $I \Rightarrow J \in \widehat{\mathbf{C}}$.

**Proof**. Suppose that $I \in \widehat{P}$, $J \in \widehat{\mathbf{C}}$, and $S \subseteq I \Rightarrow J$ with $S \in \mathbf{C}(p)$. Define $U = \{q \in I: \exists s \in S.\ q \leq s\}$. Then $U \subseteq J$. If $q \in I \cap p\!\downarrow$, then there is $T \in \mathbf{C}(q)$ for which $T \prec S$. Then for any $t \in T$, there is $s \in S$ for which $t \leq s$, whence $t \in U$. Accordingly $T \subseteq U \subseteq J$. Since $J$ is a **C**-closed, it follows that $q \in J$. We conclude that $I \cap p\!\downarrow \subseteq J$, whence $p \in p\!\downarrow \subseteq I \Rightarrow J$. Therefore $I \Rightarrow J$ is **C**-closed. □

It follows from the lemma that $\widehat{\mathbf{C}}$ is a frame. For clearly an arbitrary intersection of **C**-closed sieves is **C**-closed. So $\widehat{\mathbf{C}}$ is a complete lattice. In view of the lemma the implication operation in $\widehat{P}$ restricts to one in $\widehat{\mathbf{C}}$, making $\widehat{\mathbf{C}}$ a Heyting algebra, and so a frame.

**Proposition 1.** Suppose that **C** is a pretopology. Then **(i)** the bottom element of $\widehat{\mathbf{C}}$ is $\mathbf{0} = \{p: \varnothing \in \mathbf{C}(p)\}$, **(ii)** the **C**-closed sieve generated by a sieve $A$ (that is, the smallest **C**-closed sieve containing $A$) is $\{p: \exists S \in \mathbf{C}(p).\ S \subseteq A\}$, **(iii)** the join operation in $\widehat{\mathbf{C}}$ is given by $\bigvee_{i \in I} J_i = (\bigcup_{i \in I} J_i)^*$. If **C** is a Grothendieck topology, then **(iv)** for any sieve $S \subseteq p\!\downarrow$, $p \in S^* \leftrightarrow S \in \mathbf{C}(p)$.

**Proof.** Suppose that **C** is a pretopology. Then **0** is a **C**-closed sieve. For it is easily seen to be a sieve; and it is **C**-closed because if $S \in \mathbf{C}(p)$ and $S \subseteq \mathbf{0}$, then $\varnothing \in \mathbf{C}(s)$ for each $s \in S$, whence $\varnothing = \bigcup_{s \in S} \varnothing \in \mathbf{C}(p)$, and so $p \in \mathbf{0}$. Finally, $\mathbf{0} \subseteq I$ for any **C**-closed sieve $I$, for if $\varnothing \in \mathbf{C}(p)$, then from $\varnothing \subseteq I$ we infer $p \in I$. This gives **(i).** As for **(ii)**, suppose given a sieve $A$. Then $A \subseteq A^*$ follows from $\{p\} \in \mathbf{C}(p)$. $A^*$ is a sieve, since if $p \in A^*$ and $q \leq p$, then there is $S \in \mathbf{C}(p)$ for which $S \subseteq A$, and $T \in \mathbf{C}(q)$ sharpening $S$; clearly $T \subseteq A$ also, whence $q \in A^*$. And $A^*$ is **C**-closed, since if $S \subseteq A^*$ with $S \in$

**C**(p), then for each $s \in S$ there is $T_s \in \mathbf{C}(s)$ with $T_s \subseteq A$; it follows that $\bigcup_{s \in S} T_s \subseteq A$ and $\bigcup_{s \in S} T_s \in \mathbf{C}(p)$, whence $p \in A^*$. Part **(iii)** is an immediate consequence of **(iii)**. Finally, if **C** is a Grothendieck topology and $S \subseteq p\!\downarrow$ is a sieve, then $p \in S^* \leftrightarrow \exists T \in \mathbf{C}(p) . T \subseteq S \leftrightarrow S \in \mathbf{C}(p)$, i.e. **(iv)**. □

We observe parenthetically that $\widehat{\mathbf{Den}}$ *is a Boolean algebra*. To establish this it suffices to show that, for any $I \in \widehat{\mathbf{Den}}$, $\neg\neg I \subseteq I$. Now since $\varnothing \notin \mathbf{Den}(p)$, it follows from **(i)** of the proposition above that the bottom element of $\widehat{\mathbf{Den}}$ is $\varnothing$, so that, for any $I \in \widehat{\mathbf{Den}}$, $\neg I = \{p : I \cap p\!\downarrow = \varnothing\}$, whence $\neg\neg I = \{p : \forall q \leq p \exists r \leq q . r \in I\}$. But it easily checked that the defining condition for $I$ to be a member of $\widehat{\mathbf{Den}}$ is precisely that, if $\forall q \leq p \exists r \leq q . r \in I$, then $p \in I$. That is, $\neg\neg I \subseteq I$.

Cover schemes on $P$ correspond to certain self-maps on $\widehat{P}$ called (*weak*) *nuclei*. A *weak nucleus* on a frame $H$ is a finite-meet-preserving map $j: H \to H$ such that $j(1) = 1$ and $a \leq j(a)$ for any $a \in H$. If in addition $j(j(a)) \leq j(a)$ (so that $j(j(a)) = j(a)$) for all $a \in H$, $j$ is called a *nucleus* on $H$.

**Proposition 2.** Let **C** be a cover scheme on $P$. For each $I \in \widehat{P}$ let $I^*$ be the least **C**-closed sieve containing $I$. Then the map $k_\mathbf{C}: I \mapsto I^*$ is a nucleus on $\widehat{P}$.

**Proof.** Clearly $I \subseteq I^*$ and $I^{**} = I^*$. It remains to be shown that, for $I, J \in \widehat{P}$, $(I \cap J)^* = I^* \cap J^*$. Since $*$ is obviously inclusion-preserving, $(I \cap J)^* \subseteq I^* \cap J^*$. For the reverse inclusion, note first that $I \in \widehat{\mathbf{C}} \leftrightarrow I^* = I$. Given $I, J \in \widehat{P}$, define $K = I \Rightarrow (I \cap J)^*$. By the Lemma above, $K \in \widehat{\mathbf{C}}$, so that $K^* = K$. Now $J^* \subseteq K$ since

$$J \cap I \subseteq (I \cap J)^* \to J \subseteq [I \Rightarrow (I \cap J)^*] = K,$$

whence $J^* \subseteq K^* \subseteq K$. Similarly, if we define $L = K \Rightarrow (I \cap J)^*$, then $I^* \subseteq L$. It follows that

$$I^* \cap J^* \subseteq K \cap L = K \cap [K \Rightarrow (I \cap J)^*] \subseteq (I \cap J)^*. \quad \square$$

Inversely, any *weak* nucleus $j$ on $\widehat{P}$ determines a *regular* cover scheme $\mathbf{D}_j$ on $P$, given by

$$S \in \mathbf{D}_j(p) \leftrightarrow p \in j(S).$$

Let us check that $\mathbf{D}_j$ is indeed a regular cover scheme. To do this it suffices to show that each $\mathbf{D}_j(p)$ is a filter of sieves and that, if $S \in \mathbf{D}_j(p)$, and $q \leq p$, then $S \cap q{\downarrow} \in \mathbf{D}_j(q)$. The first of these properties follows immediately from the fact that $j$ preserves finite intersections, and the second from the observation that, if $S \in \mathbf{D}_j(p)$, and $q \leq p$, then $p \in j(S)$, so that $q \in j(S)$, and $q \in q{\downarrow} \subseteq j(q{\downarrow})$, whence $q \in j(S) \cap j(q{\downarrow}) = j(S \cap q{\downarrow})$, i.e. $S \cap q{\downarrow} \in \mathbf{D}_j(q)$.

When $j$ is a *nucleus*, $\mathbf{D}_j$ is a *Grothendieck topology*. For under this assumption, if $S \in \mathbf{D}_j(p)$ and $T_s \in \mathbf{D}_j(s)$ for each $s \in S$, then $s \in j(T_s)$ for each $s \in S$, and it follows that

$$S \subseteq \bigcup_{s \in S} j(T_s) \subseteq j\!\left(\bigcup_{s \in S} T_s\right)$$

so that

$$p \in j(S) \subseteq j\!\left(j\!\left(\bigcup_{s \in S} T_s\right)\right) = j\!\left(\bigcup_{s \in S} T_s\right)$$

i.e., $\bigcup_{s \in S} T_s \in \mathbf{D}_j(p)$.

The correspondences $\mathbf{C} \mapsto k_\mathbf{C}$ and $j \mapsto \mathbf{D}_j$ between Grothendieck topologies on $P$ and nuclei on $\widehat{P}$ are mutually inverse. For if $\mathbf{C}$ is a Grothendieck topology on $P$, then, by Proposition I **(iv)** we have

$$S \in \mathbf{D}_{k_\mathbf{C}}(p) \leftrightarrow p \in k_\mathbf{C}(S) = S^* \leftrightarrow S \in \mathbf{C}(p),$$

whence $\mathbf{D}_{k_\mathbf{C}} = \mathbf{C}$. And, for a nucleus $j$ on $\widehat{P}$, we have, using Proposition 1**(ii)**,

$$k_{\mathbf{D}_j}(I) = \text{least } \mathbf{D}_j-\text{closed sieve} \supseteq I$$
$$= \{p : \exists S \in \mathbf{D}_j(p).S \subseteq I\}$$
$$= \{p : \exists S \subseteq I.p \in j(S)\}$$
$$= j(I),$$

whence $k_{\mathbf{D}_j} = j$.

## II. COVER SCHEMES AND FRAMES

The relationship between cover schemes on a preordered set and (weak) nuclei on its completion can be extended to cover schemes on partially ordered sets and general frames. Accordingly let $H$ be a frame: we write $\bigvee$, $\wedge$, $\Rightarrow$ for the join, meet and implication operations, respectively, in $H$. The partially ordered set $(P, \leq)$ is said to be *dense* in $H$ if $P$ is a subset of $H$, the partial ordering on $P$ is the restriction to $P$ of that of $H$, and either of the two following equivalent conditions is satisfied: (i) for any $a \in H$, $a = \bigvee\{p: a \leq p\}$ (ii) for any $a, b \in H$, $a \leq b \leftrightarrow \forall p[p \leq a \to p \leq b]$. The canonical example of a frame in which $P$ is dense is the frame $\widehat{P}$ described in section I: here each $p \in P$ is identified with the $p\!\downarrow \in \widehat{P}$. $\widehat{P}$ is easily seen to have the property that in it, for any $S \subseteq P$, $p \leq \bigvee S$ iff $p \in S$.

Now fix a frame $H$ in which $P$ is dense and a cover scheme $\mathbf{C}$ on $P$. An element $a \in H$ is said to *cover* an element $p \in P$ if there exists a cover $S$ of $p$ for which $\bigvee S \leq a$. A $\mathbf{C}$-*element* of $H$ is one which dominates every element of $P$ that it covers—that is, an element $a \in H$ satisfying

$$\forall p \in P[(\exists S \in \mathbf{C}(p))\bigvee S \leq a \to p \leq a].$$

We write $H_{\mathbf{C}}$ for the set of all $\mathbf{C}$-elements of $H$. It is evident that $H_{\mathbf{C}}$ is closed under the meet operation of $H$. Notice that $\mathbf{C}$-elements and $\overline{\mathbf{C}}$-elements coincide (recalling that $\overline{\mathbf{C}}$ is the normalization of $\mathbf{C}$.)

The *canonical H-cover scheme* $\mathbf{C}_H$ on $P$ is given by

$$S \in \mathbf{C}_H(p) \leftrightarrow \bigvee S = p.$$

Clearly $\mathbf{C}_H$ is a pretopology, and every element of $H$ is a $\mathbf{C}_H$-element.

Corresponding to the Lemma of §I, we have:

**Lemma.** If $a \in H, b \in H_\mathbf{c}$, then $a \Rightarrow b \in H_\mathbf{c}$.

**Proof.** Suppose $a \in H, b \in H_\mathbf{c}$, $S \in C(p)$ and $\bigvee S \leq (a \Rightarrow b)$. Writing $U$ for $\{q : q \leq a \ \& \ \exists s \in S(q \leq s)\}$, we have

$$\bigvee U \leq \bigvee \{s \wedge q : s \in S, q \leq a\} = \bigvee S \wedge \bigvee \{q : q \leq a\} = \bigvee S \wedge a \leq b.$$

Now if $q \leq p \wedge a$, there is $T \in \mathbf{C}(q)$ sharpening $S$. Then

$$t \in T \rightarrow t \leq a \ \& \ \exists s \in S(t \leq s),$$

so that $T \subseteq U$, and therefore $\bigvee T \leq \bigvee U \leq b$. Since $b \in H_\mathbf{c}$, it follows that $q \leq b$. Hence $q \leq p \wedge a \rightarrow q \leq b$, so that $p \wedge a \leq b$ and $p \leq (a \Rightarrow b)$. We conclude that $(a \Rightarrow b) \in H_\mathbf{c}$. □

It follows from the lemma that $H_\mathbf{c}$ is itself a frame.

The nucleus on $H$ *associated* with the cover scheme $\mathbf{C}$ on $P$ is the map $j = k_\mathbf{C} \colon H \rightarrow H$ defined by

$$j(a) = \bigwedge \{x \in H_\mathbf{c} : a \leq x\}.$$

That $j$ is a nucleus results from the following observations. Evidently $j$ is order preserving, maps $H$ onto $H_\mathbf{c}$, is the identity on $H_\mathbf{c}$, and satisfies $j(1) = 1$ and $a \leq j(a)$ for all $a \in A$. Also it is easily shown that $j(j(a)) = j(a)$. Finally, $j$ preserves finite meets. For clearly $j(a \wedge b) \leq j(a) \wedge j(b)$ since $j$ is order preserving. For the reverse inequality, consider first the element $u = (a \Rightarrow j(a \wedge b))$: this is, by the Lemma above, an element of $H_\mathbf{c}$, so that $j(u) = u$. Also $j(b) \leq u$. For from $b \wedge a \leq j(a \wedge b)$ we deduce $b \leq (a \Rightarrow j(a \wedge b)) = u$, whence $j(b) \leq j(u) = u$. Similarly, $v = ((a \Rightarrow j(a \wedge b)) \Rightarrow j(a \wedge b))$ is an element of $H_C$ and $j(a) \leq v$. Therefore

$$j(a) \wedge j(b) \leq v \wedge u \leq j(a \wedge b),$$

as required.

Notice that the nucleus associated with a cover scheme coincides with that associated with its normalization.

Accordingly we have shown that each cover scheme on $P$ determines a nucleus on $H$. Conversely, we can show that any *weak nucleus* on $H$ determines a cover scheme on $P$. For, starting with a weak nucleus $j$ on $H$, define the map $\mathbf{D}_j$ on $P$ by

$$\mathbf{D}_j(p) = \{S \subseteq p\downarrow : p \leq j(\bigvee S)\}.$$

Then $\mathbf{D}_j$ is a cover scheme on $P$. For suppose $q \leq p$ and $S \in \mathbf{D}_j(p)$. Then $q \leq p \leq j(\bigvee S)$; since $q \leq j(q)$ and $j$ preserves finite meets, it follows that

(*) $\qquad q \leq j(q) \wedge j(\bigvee S) = j(q \wedge \bigvee S) = j(\bigvee\{s \wedge q : s \in S\}).$

Now define $T \subseteq q\downarrow$ by

$$T = \{t : t \leq q \ \&\ \exists s \in S(t \leq s)\}.$$

We claim that $T$ is a ($\mathbf{D}_j$-) cover of $q$ sharpening $S$. That $T$ sharpens $S$ is evident from its definition. To see that it is a cover of $q$ we observe that, if $s \in S$, then

$$s \wedge q = \bigvee\{t : t \leq s \wedge q\} = \bigvee\{t : t \leq s\ \&\ t \leq q\} \leq \bigvee T.$$

Therefore $\bigvee\{s \wedge q : s \in S\} \leq \bigvee T$, so that, by (*),
$$q \leq j(\bigvee\{s \wedge q : s \in S\}) \leq j(\bigvee T),$$
that is, $T$ covers $q$.

When $j$ is a *nucleus,* the associated cover scheme $\mathbf{D}_j$ is actually a *pretopology*. For in any case $\{p\} \in \mathbf{D}_j(p)$. Moreover, if $j$ is a nucleus, $S \in \mathbf{D}_j(p)$ and $T_s \in \mathbf{D}_j(s)$ for each $s \in S$, then

$$p \leq j(\bigvee S) \leq j(\bigvee_{s \in S} j(\bigvee T_s)) \leq j(j(\bigvee_{s \in S} \bigvee T_s)) = j(\bigvee \bigcup_{s \in S} T_s).$$

Therefore $\bigcup_{s \in S} T_s \in \mathbf{D}_j(p)$, and $\mathbf{D}_j$ is a pretopology.

Starting with a weak nucleus $j$, we obtain the corresponding cover scheme $\mathbf{D}_j$. The latter in turn determines a nucleus $j^*$ given by

$$j^*(a) = \bigwedge\{x \in H_{\mathbf{D}_j} : a \leq x\}.$$

Now by definition, we have

$$x \in H_{\mathbf{D}_j} \leftrightarrow \forall p[(\exists S \in \mathbf{D}_j(p)) \bigvee S \leq x \to p \leq x]$$
$$\leftrightarrow \forall p[(\exists S \subseteq p\downarrow)(p \leq j(\bigvee S) \,\&\, \bigvee S \leq x) \to p \leq x] \quad \text{(a)}$$
$$\leftrightarrow \forall p[p \leq j(x) \to p \leq x] \quad \text{(b)}$$
$$\leftrightarrow j(x) = x.$$

(To see the equivalence between (a) and (b), we need to establish the equivalence between $(\exists S \subseteq p\downarrow)(p \leq j(\bigvee S) \,\&\, \bigvee S \leq x)$ and $p \leq j(x)$. Clearly the first of these implies the second. As for the converse, if $p \leq j(x)$, then since $p \leq j(p)$, we have

$$p \leq j(x) \wedge j(p) = j(x \wedge p) = j(\bigvee S),$$

where $S = \{q : q \leq x \wedge p\}$. Then $S \subseteq p\downarrow$, $p \leq j(\bigvee S)$ and $\bigvee S \leq x$, and the first statement follows.) Accordingly

(*) $\qquad\qquad j^*(a) = \bigwedge\{x \in H : a \leq x \,\&\, j(x) = x\}.$

$j^*$ is called the nucleus[2] *generated* by the weak nucleus $j$; it is easily deduced from (*) that when $j$ is a nucleus, $j^*$ and $j$ coincide.

The generation of nuclei by weak nuclei can itself be seen as an instance of a nuclear operation. For consider the set $W(H)$ of all weak nuclei on $H$. When $W(H)$ is partially ordered pointwise in the obvious way, it becomes a frame with implications, joins, and meets given by the following specifications: $(j \Rightarrow k)(a) = \bigwedge_{b \geq a}(j(b) \Rightarrow k(b))$ and for $S \subseteq W(H)$,

$(\bigvee S)(a) = \bigvee_{s \in S} s(a)$, $(\bigwedge S)(a) = \bigwedge_{s \in S} s(a)$. The subset $N(H)$ of $W(H)$ consisting of all nuclei can be shown to be a sublocale (see Johnstone [1]) of $N(H)$, that is, it is closed under arbitrary meets in $W(H)$ and is such that $(j \Rightarrow k) \in N(H)$ whenever $j \in W(H)$, $k \in N(H)$. That being the case, the map $\varphi: W(H) \to N(H)$ defined by

$$\varphi(j) = \bigwedge\{k \in N(H) : j \leq k\}$$

---

[2] It can be verified directly that $j^*$ is a nucleus.

is a nucleus on $W(H)$, and it is easily shown that $\varphi(j) = j^*$. So the generation of nuclei by weak nuclei is precisely the action of the nucleus $\varphi$.

Now start with a cover scheme $\mathbf{C}$ on $P$, obtain the associated nucleus $k_\mathbf{C}$ on $H$, and consider its associated cover scheme $\mathbf{D}_{k_\mathbf{C}} = \mathbf{C}^*$ on $P$. By definition we have, for $S \subseteq p\!\downarrow$,

$$\begin{aligned} S \in \mathbf{C}^*(p) &\leftrightarrow p \leq j_\mathbf{C}(\bigvee S) \\ &\leftrightarrow p \leq \bigwedge\{x \in H_\mathbf{C} : \bigvee S \leq x\} \\ &\leftrightarrow \forall x \in H_\mathbf{C}(\bigvee S \leq x \to p \leq x) \end{aligned}$$

Recalling the definition of $H_\mathbf{C}$, we see immediately that this last assertion is implied by $S \in \mathbf{C}(p)$, so that always $\mathbf{C}(p) \subseteq \mathbf{C}^*(p)$. The reverse inclusion will hold, and so $\mathbf{C}$ will coincide with $\mathbf{C}^*$, precisely when the cover scheme $\mathbf{C}$ is *saturated*, that is, coincides with its *saturate*, which we next proceed to define.

The ($H$-) *saturate* $\widetilde{\mathbf{C}}$ of a cover scheme $C$ on $P$ is defined by

$$\widetilde{\mathbf{C}}(p) = \{S \subseteq p\!\downarrow : \forall x \in H_\mathbf{C}(\bigvee S \leq x \to p \leq x)\}.$$

Then $\widetilde{\mathbf{C}}$ is a cover scheme. For if $S \in \widetilde{\mathbf{C}}(p)$ and $q \leq p$, consider the subset $T$ of $p\!\downarrow$ defined by

$$T = \{t \leq q : \exists s \in S(t \leq s)\}.$$

It is easily shown that $\bigvee T = (\bigvee S) \wedge q$. Now if $x \in H_\mathbf{C}$ and $\bigvee T \leq x$, then $\bigvee S \wedge q \leq x$, whence $\bigvee S \leq (q \Rightarrow x)$. But since $x$ is an element of $H_\mathbf{C}$, so, by the lemma, is $q \Rightarrow x$, and since $S \in \widetilde{\mathbf{C}}(p)$, it follows that $p \leq (q \Rightarrow x)$. Thus $q = p \wedge q \leq x$. Accordingly $T \in \widetilde{\mathbf{C}}(q)$, and $T$ obviously sharpens $S$. This shows that $\widetilde{\mathbf{C}}$ is indeed a cover scheme.

It is readily shown that any cover scheme associated with a nucleus (as opposed to a weak nucleus) is saturated. Observe that, when $H$ is $\widehat{P}$, every coverage on $P$ is saturated, since in that case $H_\mathbf{C}$ is $\widehat{\mathbf{C}}$ and so we have, using Proposition I.1 (**iv**),

$$S \in \widetilde{\mathbf{C}}(p) \leftrightarrow \forall I \in \widehat{\mathbf{C}}[S \subseteq I \to p \in I] \leftrightarrow p \in S^* \leftrightarrow S \in \mathbf{C}(p).$$

To sum up, each weak nucleus on $H$ gives rise to a cover scheme on $P$ and the cover scheme associated with a nucleus is saturated. Conversely, each cover scheme gives rise to a nucleus. This establishes mutually inverse correspondences between nuclei and saturated cover schemes.

Given $a \in H$, we define the nuclei $j_a$, $j^a$ on $H$ by

$$j_a(x) = a \vee x \qquad j^a(x) = a \Rightarrow x.$$

The associated cover schemes (easily seen to be pretopologies) on $P$ are given by:

$$S \in \mathbf{C}_a(p) \leftrightarrow p \leq a \vee \bigvee S$$
$$S \in \mathbf{C}^a(p) \leftrightarrow p \wedge a \leq \bigvee S.$$

Notice that

$$p \leq a \leftrightarrow \emptyset \in \mathbf{C}_a(p)$$
$$p \leq \neg a \leftrightarrow \emptyset \in \mathbf{C}^a(p).$$

The *double negation operation* $\neg\neg$ is a nucleus on $H$, whose associated cover scheme is precisely the dense cover scheme **Den** (which accordingly is also known as the *double negation cover scheme*). An argument similar to the one above showing that $\widehat{\mathbf{Den}}$ is a Boolean algebra establishes that $H_{\mathbf{Den}}$ is a Boolean algebra: it is in fact the complete Boolean algebra of $\neg\neg$-closed elements of $H$.

**Proposition.** Let $j$ be a weak nucleus on $H$. Then the following are equivalent: (a) $j(0) = 0$ (b) $j \leq \neg\neg$ (in the pointwise ordering of $W(H)$) (c) $\emptyset \notin \mathbf{D}_j(p)$ for all $p$.

**Proof.** If $j \leq \neg\neg$ then $j0 \leq \neg\neg 0 = 0$. Conversely if $j0 = 0$ then, for any $a \in H$, $j(a) \wedge \neg a \leq j(a) \wedge j(\neg a) = j(a \wedge \neg a) = j0 = 0$. So $j(a) \leq \neg\neg a$. Finally, we have

$$j0 = 0 \leftrightarrow 0 \in H_{\mathbf{D}_j} \leftrightarrow \forall p[(\exists S \in \mathbf{D}_j(p)) \bigvee S = 0 \to p = 0]$$
$$\leftrightarrow \forall p[\neg(\exists S \in \mathbf{D}_j(p) \bigvee S = 0]$$
$$\leftrightarrow \forall p[\emptyset \notin \mathbf{D}_j(p)]. \quad \square$$

### III. COVER SCHEMES AND KRIPKE MODELS

We revert to the assumption that $(P, \leq)$ is a preordered set. Recall that a *presheaf* on $P$ is an assignment, to each $p \in P$, of a set $\mathscr{F}(p)$ and to each pair $(p, q)$ with $q \leq p$ of a map $\mathscr{F}_{pq}\colon \mathscr{F}(p) \to \mathscr{F}(q)$ in such a way that $\mathscr{F}_{pp}$ is the identity on $\mathscr{F}(p)$ and, for $r \leq q \leq p$, $\mathscr{F}_{pr} = \mathscr{F}_{qr} \circ \mathscr{F}_{pq}$. The set $V(\mathscr{F}) = \bigcup_{p \in P} \mathscr{F}(p)$ is called the *universe* of $\mathscr{F}$. A *Kripke model* based on $P$ is a presheaf $\mathscr{K}$ for which $\mathscr{K}(p) \subseteq \mathscr{K}(q)$ whenever $q \leq p$ and each $\mathscr{K}_{pq}$ is the corresponding insertion map. Put more simply, a Kripke model based on $P$ is a map $\mathscr{K}$ from $P$ to a family of sets satisfying $\mathscr{K}(p) \subseteq \mathscr{K}(q)$ whenever $q \leq p$. A Kripke model $\mathscr{K}$ based on $P$ may be regarded as a set "evolving" or "growing" over $P$: each $\mathscr{K}(p)$ may be thought of as the "state" of the evolving set $\mathscr{K}$ at "stage" $p$.

Now suppose that we are given a cover scheme $\mathbf{C}$ on $P$. A Kripke model $\mathscr{K}$ based on $P$ satisfying

$$\mathscr{K}(p) = \bigcap_{s \in S} \mathscr{K}(s)$$

for any $p \in P$, $S \in \mathbf{C}(p)$ is said to be *compatible with* $\mathbf{C}$. (When $P$ is directed downward, that is, whenever each pair of elements of $P$ has a lower bound, and $\mathbf{C}$ is a pretopology on $P$, a Kripke model compatible with $\mathbf{C}$ is nothing other than a $\mathbf{C}$-*sheaf*.)

Each Kripke model $\mathcal{K}$ based on $P$ induces a Kripke model $\widetilde{\mathcal{K}}$ based on the free lower semilattice $\widetilde{P}$ generated by $P$ by setting

$$\widetilde{\mathcal{K}}(\varnothing) = \varnothing \qquad \widetilde{\mathcal{K}}(\{p_1,\ldots,p_n\}) = \mathcal{K}(p_1) \cup \ldots \cup \mathcal{K}(p_n).$$

If, further, $\mathcal{K}$ is compatible with the cover scheme **C** on $P$, then $\widetilde{\mathcal{K}}$ is compatible with the cover scheme on $\widetilde{P}$ induced by $P$ (and hence also with the associated coverage on $\widetilde{P}$.) For suppose that $\mathcal{K}$ is in fact compatible with the cover scheme **C** on $P$. Given $F \in \widetilde{P}$, a nonempty subset $\{p_1,\ldots,p_n\}$ of $F$, and $S_1 \in \mathbf{C}(p_1),\ldots,S_n \in \mathbf{C}(p_n)$, we have

$$\bigcap_{X \in S_1 \bullet \ldots \bullet S_n} \widetilde{\mathcal{K}}(X) = \bigcap_{s_1 \in S_1,\ldots,s_n \in S_n} \widetilde{\mathcal{K}}(\{s_1,\ldots,s_n\} \cup F)$$

$$= \bigcap_{s_1 \in S_1,\ldots,s_n \in S_n} (\mathcal{K}(s_1) \cup \ldots \cup \mathcal{K}(s_n) \cup \bigcup_{p \in F} \mathcal{K}(p))$$

$$= \bigcap_{s_1 \in S_1} \mathcal{K}(s_1) \cup \ldots \cup \bigcap_{s_n \in S_n} \mathcal{K}(s_n) \cup \bigcup_{p \in F} \mathcal{K}(p))$$

$$= \mathcal{K}(p_1) \cup \ldots \cup \mathcal{K}(p_n) \cup \bigcup_{p \in F} \mathcal{K}(p)$$

$$= \widetilde{\mathcal{K}}(F).$$

Now suppose that the cover scheme **C** is in fact a *pretopology*. Then any Kripke model $\mathcal{K}$ based on $P$ induces a Kripke model $\mathcal{K}_\mathbf{C}$ also based on $P$ but in addition compatible with **C** given by

$$\mathcal{K}_\mathbf{C}(p) = \bigcup_{S \in \mathbf{C}(p)} \bigcap_{s \in S} \mathcal{K}(s),$$

that is,

$$a \in \mathcal{K}_\mathbf{C}(p) \leftrightarrow \exists S \in \mathbf{C}(p) \forall s \in S.\ a \in \mathcal{K}(s).$$

We note that $\mathcal{K}(p) \subseteq \mathcal{K}_\mathbf{C}(p)$ for every $p$. It is easily checked that this defines a Kripke model over $P$; let us confirm its compatibility with **C**. It suffices to show that, given $S \in \mathbf{C}(p)$, we have $\bigcap_{s \in S} \mathcal{K}_\mathbf{C}(s) \subseteq \mathcal{K}_\mathbf{C}(p)$. Indeed, if $a \in \bigcap_{s \in S} \mathcal{K}_{\mathbf{C}}(s)$, then for each $s \in S$ there is $T_s \in \mathbf{C}(s)$ with $a \in \bigcap_{t \in T_s} \mathcal{K}(t)$. Writing $T = \bigcup_{s \in S} T_s$, we then have $T \in \mathbf{C}(p)$ and $a \in \bigcap_{t \in T} \mathcal{K}(t)$. It follows that $a \in \mathcal{K}_{\mathbf{C}}(p)$, as required.

Let us now examine some special cases. Let $U$ be a subset of the universe $V$ of $\mathcal{K}$, and let $U^*$ be the sieve $\{p: U \subseteq \mathcal{K}(p)\}$. Now consider the Kripke model $\mathcal{K}^U$ compatible with $\mathbf{C}^{U^*}$ induced by $\mathcal{K}$. For arbitrary $p \in P$, we have $S = p\!\downarrow \cap\, U^* \in \mathbf{C}^{U^*}(p)$. and $U \subseteq \mathcal{K}(s) \subseteq \mathcal{K}^U(s)$ for every $s \in S$. Hence $U \subseteq \bigcap_{s \in S} \mathcal{K}^U(s) = \mathcal{K}^U(p)$. Thus, under these conditions, $U$ is a subset of every $\mathcal{K}^U(p)$. In other words, the passage from $\mathcal{K}$ to $\mathcal{K}^U$ *forces* $U$ to be included in the state of $\mathcal{K}^U$ at each stage. (Note that if $U^* = \emptyset$ then $\mathcal{K}^U$ assumes the constant value $V$.)

Again let $U$ be a subset of $V$; this time define $U^+$ to be the sieve $\{p: U \cap \mathcal{K}(p) \neq \emptyset\}$. Now consider the Kripke model $\mathcal{K}_U$ compatible with $\mathbf{C}_{U^+}$ induced by $\mathcal{K}$. Then for any $p$ we have

$$U \cap \mathcal{K}(p) \neq \emptyset \to p \in U^+ \to \emptyset \in \mathbf{C}_{U^+}(p) \to \mathcal{K}_U(p) = \bigcap_{s \in \emptyset} \mathcal{K}_U(s) = V.$$

That is, the passage from $\mathcal{K}$ to $\mathcal{K}_U$ *forces* each state of $\mathcal{K}_U$, apart from those already maximal, to be disjoint from $U$. (Notice that if $U^+ = P$, then $\mathcal{K}_U$ assumes the constant value $V$.)

We turn turn to *logic* in Kripke models. Each Kripke model $\mathcal{K}$ based on $P$, with universe $V$, determines a map $\widehat{\mathcal{K}}: V \to \widehat{P}$ given by

$$\widehat{\mathcal{K}}(v) = \{p : v \in \mathcal{K}(p)\}.$$

This extends naturally to a homomorphism—also denoted by $\widehat{\mathcal{K}}$—of the free Heyting algebra $F(V)$ generated by $V$ into $\widehat{P}$. Think of the members of $F(V)$ as the formulas of intuitionistic propositional logic generated by the members of $V$ regarded as propositional atoms. Introduce the familiar *forcing* relation $\Vdash_{\mathcal{K}}$ between $P$ and $F(V)$ by defining

(*) $$p \Vdash_{\mathcal{K}} \varphi \leftrightarrow p \in \widehat{\mathcal{K}}(\varphi).$$

Then the fact that $\widehat{\mathcal{K}}: F(V) \to \widehat{P}$ is a homomorphism of Heyting algebras translates into the usual rules for "Kripke semantics", namely

- $p \Vdash_{\mathcal{K}} \varphi \wedge \psi \leftrightarrow p \Vdash_{\mathcal{K}} \varphi \ \& \ p \Vdash_{\mathcal{K}} \psi$
- $p \Vdash_{\mathcal{K}} \varphi \vee \psi \leftrightarrow p \Vdash_{\mathcal{K}} \varphi \ \text{or} \ p \Vdash_{\mathcal{K}} \psi$
- $p \Vdash_{\mathcal{K}} \varphi \Rightarrow \psi \leftrightarrow \forall q \leq p [\, q \Vdash_{\mathcal{K}} \varphi \to q \Vdash_{\mathcal{K}} \psi\,]$
- $p \Vdash_{\mathcal{K}} \neg \varphi \leftrightarrow \forall q \leq p \ q \nVdash_{\mathcal{K}} \varphi$

Equally, the map $\widehat{\mathcal{K}}: V \to \widehat{P}$ extends to a frame homomorphism (i.e., a map preserving top elements, $\wedge$, and $\bigvee$)—again denoted by $\widehat{\mathcal{K}}$—of the *free frame* $\Phi(V)$ generated by $V$. Think of the members of $\Phi(V)$ as the formulas of *infinitary* intuitionistic propositional logic generated by the members of $V$ regarded as propositional atoms. Such a formula $\varphi$ is said to be *geometric* if it is generated from propositional atoms by applying just $\wedge$ and $\bigvee$. Introducing the forcing relation $\Vdash_{\mathcal{K}}$ between $P$ and $\Phi(V)$ as in (*) above, the fact that $\widehat{\mathcal{K}}: \Phi(V) \to \widehat{P}$ is a frame homomorphism translates into the semantical rules for *geometric* formulas:

- $p \Vdash_{\mathcal{K}} \varphi \wedge \psi \leftrightarrow p \Vdash_{\mathcal{K}} \varphi \ \& \ p \Vdash_{\mathcal{K}} \psi$
- $p \Vdash_{\mathcal{K}} \bigvee_{i \in I} \varphi_i \leftrightarrow \exists i \in I \ p \Vdash_{\mathcal{K}} \varphi_i$.

Now suppose that $\mathbf{C}$ is a pretopology on $P$. It is then easily seen that $\mathcal{K}$ is compatible with $\mathbf{C}$ iff each $\widehat{\mathcal{K}}(v)$ is a $\mathbf{C}$-closed sieve. So if $\mathcal{K}$ is compatible with $\mathbf{C}$, the resulting map $\widehat{\mathcal{K}}: V \to \widehat{\mathbf{C}}$ can be extended to a homomorphism, which we shall denote by $\widehat{\mathcal{K}}_{\mathbf{C}}$, of $F(V)$ into $\widehat{\mathbf{C}}$. Introducing the forcing relation $\Vdash_{\mathcal{K},\mathbf{C}}$ between $P$ and $F(V)$ by

(**) $$p \Vdash_{\mathcal{K},\mathbf{C}} \varphi \leftrightarrow p \in \widehat{\mathcal{K}_\mathbf{C}}(\varphi),$$

we find that the fact that $\widehat{\mathcal{K}_\mathbf{C}}: F(V) \to \widehat{\mathbf{C}}$ is a homomorphism translates into the rules of "Beth-Kripke-Joyal" semantics for $\Vdash_{\mathcal{K},\mathbf{C}}$ (see, e.g., [4]), viz.,

- $p \Vdash_{\mathcal{K},\mathbf{C}} \varphi \wedge \psi \leftrightarrow p \Vdash_{\mathcal{K},\mathbf{C}} \varphi \ \& \ p \Vdash_{\mathcal{K},\mathbf{C}} \psi$
- $p \Vdash_{\mathcal{K},\mathbf{C}} \varphi \vee \psi \leftrightarrow \exists S \in \mathbf{C}(p) \ \forall s \in S \ [s \Vdash_{\mathcal{K},\mathbf{C}} \varphi \ \text{or} \ s \Vdash_{\mathcal{K},\mathbf{C}} \psi]$
- $p \Vdash_{\mathcal{K},\mathbf{C}} \varphi \Rightarrow \psi \leftrightarrow \forall q \leq p [\ q \Vdash_{\mathcal{K},\mathbf{C}} \varphi \ \to \ q \Vdash_{\mathcal{K},\mathbf{C}} \psi]$
- $p \Vdash_{\mathcal{K},\mathbf{C}} \neg \varphi \leftrightarrow \forall q \leq p \ [q \Vdash_{\mathcal{K},\mathbf{C}} \varphi \to \varnothing \in \mathbf{C}(p)]$.

We verify the second and fourth of these. We have, using Proposition I 1. (iii),

$$p \Vdash_{\mathcal{K},\mathbf{C}} \varphi \vee \psi \leftrightarrow p \in \widehat{\mathcal{K}_\mathbf{C}}(\varphi \vee \psi) = \widehat{\mathcal{K}_\mathbf{C}}(\varphi) \vee \widehat{\mathcal{K}_\mathbf{C}}(\psi) \ (\text{in } \widehat{\mathbf{C}})$$

$$\leftrightarrow \exists S \in \mathbf{C}(p). \ S \subseteq \widehat{\mathcal{K}_\mathbf{C}}(\varphi) \cup \widehat{\mathcal{K}_\mathbf{C}}(\psi)$$

$$\leftrightarrow \exists S \in \mathbf{C}(p) \forall s \in S. \ s \in \widehat{\mathcal{K}_\mathbf{C}}(\varphi) \vee s \in \widehat{\mathcal{K}_\mathbf{C}}(\psi)$$

$$\leftrightarrow \exists S \in \mathbf{C}(p) \ \forall s \in S \ [s \Vdash_{\mathcal{K},\mathbf{C}} \varphi \ \text{or} \ s \Vdash_{\mathcal{K},\mathbf{C}} \psi].$$

Also, using Proposition I 1. (ii) we have

$$p \Vdash_{\mathcal{K},\mathbf{C}} \neg \varphi \leftrightarrow p \in \widehat{\mathcal{K}_\mathbf{C}}(\neg \varphi) = \neg \widehat{\mathcal{K}_\mathbf{C}}(\varphi) = (\widehat{\mathcal{K}_\mathbf{C}}(\varphi) \Rightarrow \mathbf{0})$$

$$\leftrightarrow \forall q \leq p [q \in \widehat{\mathcal{K}_\mathbf{C}}(\varphi) \to \varnothing \in \mathbf{C}(q)]$$

$$\leftrightarrow \forall q \leq p [q \Vdash_{\mathcal{K},\mathbf{C}} \varphi \to \varnothing \in \mathbf{C}(q)].$$

Since $\widehat{\mathbf{Den}}$ is a Boolean algebra it follows that, when $\mathcal{K}$ is compatible with $\mathbf{Den}$, $p \Vdash_{\mathcal{K},\mathbf{Den}} \varphi \vee \neg \varphi$ for every $p$, i.e., classical logic prevails in the Kripke model associated with $\widehat{\mathcal{K}_{\mathbf{Den}}}$.

When $\mathscr{K}$ is compatible with **C**, the map $\widehat{\mathscr{K}}: V \to \widehat{\mathbf{C}}$ can be extended to a frame homomorphism, which we shall again denote by $\widehat{\mathscr{K}}_\mathbf{C}$, of $\Phi(V)$ into $\widehat{\mathbf{C}}$. Introduce the forcing relation $\Vdash_{\mathscr{K},\mathbf{C}}$, now between $P$ and $\Phi(V)$, by the same equivalence (**) as above. When **C** is a Grothendieck topology, a straightforward inductive argument shows that, for any geometric formula $\varphi$,

(†) $\qquad p \Vdash_{\mathscr{K},\mathbf{C}} \varphi \;\leftrightarrow\; \exists S \in \mathbf{C}(p) \; \forall s \in S \,.\, s \Vdash_\mathscr{K} \varphi.$

This may be applied to "force" any given set $\Sigma$ of geometric formulas to become true in a Kripke model. For, starting with a Kripke model $\mathscr{K}$, let $A$ be the sieve $\{p: \forall \sigma \in \Sigma.\; p \Vdash_\mathscr{K} \sigma\}$. Let **G** be the Grothendieck topology generated by the coverage $\mathbf{C}^A$: it is easily verified that a sieve $S \subseteq p{\downarrow}$ satisfies the same condition for membership in $\mathbf{G}(p)$ as in $\mathbf{C}^A(p)$, viz., $p{\downarrow} \cap A \subseteq S$. Now by (†) we have, for each $\sigma \in \Sigma$,

(‡) $\qquad p \Vdash_{\mathscr{K},\mathbf{G}} \sigma \;\leftrightarrow\; \exists S \in \mathbf{G}(p) \; \forall s \in S \,.\, s \Vdash_\mathscr{K} \sigma.$

If we take $S$ to be $p{\downarrow} \cap A$, then evidently $S \in \mathbf{G}(p)$ and $\forall s \in S.\; s \Vdash_\mathscr{K} \sigma$. It now follows from (‡) that $p \Vdash_{\mathscr{K},\mathbf{G}} \sigma$ for every $\sigma \in \Sigma$ and every $p \in P$. In this sense **G** "forces" all the members of $\Sigma$ to be true in the Kripke model associated with $\widehat{\mathscr{K}}_\mathbf{G}$.

### IV. COVER SCHEMES AND FRAME-VALUED SET THEORY

We now set about relating what has been done so far to frame-valued set theory. Associated with each frame $H$ is an $H$-valued model $V^{(H)}$ of (intuitionistic) set theory (see, e.g. [1] or [2]): we recall some of its principal features.

- Each of the members of $V^{(H)}$—the *H-sets*—is a map from a subset of $V^{(H)}$ to $H$.

- Corresponding to each sentence $\sigma$ of the language of set theory (with names for all elements of $V^{(H)}$) is an element $[\![\sigma]\!] = [\![\sigma]\!]^H \in H$ called its *truth value in* $V^{(H)}$. These "truth values" satisfy the following conditions. For $a, b \in V^{(H)}$,

$$[\![b \in a]\!] = \bigvee_{c \in dom(a)} [\![b = c]\!] \wedge a(c)$$

$$[\![b = a]\!] = \bigvee_{c \in dom(a) \cup dom(b)} ([\![c \in b]\!] \Leftrightarrow [\![c \in a]\!])$$

$$[\![\sigma \wedge \tau]\!] = [\![\sigma]\!] \wedge [\![\tau]\!], \text{ etc.}$$

$$[\![\exists x \varphi(x)]\!] = \bigvee_{a \in V^{(H)}} [\![\varphi(a)]\!]$$

$$[\![\forall x \varphi(x)]\!] = \bigwedge_{a \in V^{(H)}} [\![\varphi(a)]\!]$$

A sentence $\sigma$ is *valid*, or *holds*, in $V^{(H)}$, written $V^{(H)} \vDash \sigma$, if $[\![\sigma]\!] = 1$, the top element of $H$.

- The axioms of intuitionistic Zermelo-Fraenkel set theory are valid in $V^{(H)}$. Accordingly the category $\mathscr{S}et^{(H)}$ of sets constructed within $V^{(H)}$ is a topos: in fact $\mathscr{S}et^{(H)}$ can be shown to be equivalent to the topos of canonical sheaves on $H$.

- There is a canonical embedding $x \mapsto \hat{x}$ of the universe $V$ of sets into $V^{(H)}$ satisfying

$$[\![u \in \hat{x}]\!] = \bigvee_{y \in x} [\![u = \hat{y}]\!] \text{ for } x \in V, u \in V^{(H)}$$

$$x \in y \leftrightarrow V^{(H)} \vDash \hat{x} \in \hat{y}, \quad x = y \leftrightarrow V^{(H)} \vDash \hat{x} = \hat{y} \text{ for } x, y \in V$$

$$\varphi(x_1, \ldots, x_n) \leftrightarrow V^{(H)} \vDash \varphi(\hat{x}_1, \ldots, \hat{x}_n) \text{ for } x_1, \ldots, x_n \in V \text{ and restricted } \varphi$$

(Here a formula $\varphi$ is *restricted* if all its quantifiers are restricted, i.e. can be put in the form $\forall x \in y$ or $\exists x \in y$.)

It follows from the last of these assertions that the canonical representative $\widehat{H}$ of $H$ is a Heyting algebra in $V^{(H)}$. The *canonical prime filter* in $\widehat{H}$ is the $H$-set $\Phi_H$ defined by

$$dom(\Phi_H) = \{\hat{a} : a \in H\}, \quad \Phi_H(\hat{a}) = a \text{ for } a \in H.$$

Clearly $V^{(H)} \vDash \Phi_H \subseteq \widehat{H}$, and it is easily verified that

$$V^{(H)} \vDash \Phi_H \text{ is a (proper) prime filter}[3] \text{ in } \widehat{H}.$$

It can also be shown that $\Phi_H$ is *V-generic* in the sense that, for any subset $A \subseteq H$,

$$V^{(H)} \vDash \widehat{\bigvee A} \in \Phi_H \leftrightarrow \Phi_H \cap \hat{A} \neq \emptyset.$$

Moreover, for any $a \in H$ we have $[\![\hat{a} \in \Phi_H]\!] = a$, and in particular, for any sentence $\sigma$, $[\![\sigma]\!] = [\![\widehat{[\![\sigma]\!]} \in \Phi_H]\!]$. Thus $V^{(H)} \vDash \sigma \leftrightarrow V^{(H)} \vDash \widehat{[\![\sigma]\!]} \in \Phi_H$—in this sense $\Phi_H$ is the filter of "true" sentences in $V^{(H)}$.

This suggests that we define a *truth set* in $V^{(H)}$ to be an $H$-set $F$ for which

$$V^{(H)} \vDash F \text{ is a filter in } \widehat{H} \text{ such that } F \supseteq \Phi_H.$$

There is a natural bijective correspondence between truth sets in $V^{(H)}$ and weak nuclei on $H$. With each weak nucleus $j$ on $H$ we associate the $H$-set $T_j$ defined by $dom(T_j) = dom(\Phi_H)$ and $T_j(\hat{a}) = j(a)$ for $a \in H$. It is easily verified that $T_j$ is a truth set—the requirement that $T_j$ be a filter corresponds exactly to the condition that $j$ preserve finite meets and that it contain $\Phi_H$ to the condition that $j$ satisfy $a \leq j(a)$. Inversely, given a truth set $F$ in $V^{(H)}$, we define the map $j_F : H \to H$ by $j_F(a) = [\![\hat{a} \in T]\!]$. Again, it is readily verified that $j_F$ is a weak nucleus on $H$. These correspondences are evidently mutually inverse and in fact establish an isomorphism between the frame $W(H)$ of weak nuclei on $H$ and the internal frame of filters in $\widehat{H}$ containing $\Phi_H$. Under this isomorphism *nuclei* correspond precisely to *reflexive truth sets,* that is, truth sets satisfying the additional condition (evidently met by $\Phi_H$)

$$V^{(H)} \vDash \widehat{[\![\hat{a} \in F]\!]} \in F \to \hat{a} \in F.$$

---

[3] We recall that a filter $F$ in a lattice is *prime* if $x \vee y \in F$ implies $x \in F$ or $y \in F$.

It is of interest to examine the familiar case in which $H$ is a complete *Boolean algebra B*. In this case the canonical prime filter $\Phi_B$ is an *ultrafilter* in $\widehat{B}$, so that, in $V^{(B)}$, the only filters in $\widehat{B}$ containing $\Phi_B$—the only truth sets—are $\Phi_B$ itself and $\widehat{B}$. It follows that, for truth sets $F$ and $G$ in $V^{(B)}$

$$V^{(B)} \vDash F = G \leftrightarrow [\hat{0} \in F \leftrightarrow \hat{0} \in G].$$

In other words, the truth value $[\![\hat{0} \in F]\!]$, which can be an arbitrary member of $B$, determines the identity of $F$. This means that truth sets in $V^{(B)}$, and so equally weak nuclei on $B$, are in bijective correspondence with the members of $B$. In fact it is readily shown directly that any weak nucleus on a Boolean algebra $B$ is of the form $j_a$ for some $a \in B$. For given a weak nucleus $j$ on $B$, observe: $\neg x \leq j(\neg x)$, whence $\neg j(\neg x) \leq \neg\neg x = x$. Also $j(x) \wedge j(\neg x) = j(x \wedge \neg x) = j(0)$, whence $j(x) \leq j(\neg x) \Rightarrow j(0) = \neg j(\neg x) \vee j(0) \leq x \vee j(0)$. But clearly $x \vee j(0) \leq j(x)$, so that $j(x) = x \vee j(0)$.

Consider now the special case in which $H$ is the completion $\widehat{P}$ of a preordered set $P$. We have already established a bijective correspondence between Grothendieck topologies on $P$ and nuclei on $\widehat{P}$. This leads in turn to a bijective correspondence between Grothendieck topologies on $P$ and reflexive truth sets in $V^{(\widehat{P})}$. Explicitly, this correspondence assigns to each Grothendieck topology **C** on $P$ the reflexive truth set $T_\mathbf{C}$ in $V^{(\widehat{P})}$ given by $T_\mathbf{C}(S) = S^*$ for $S \in \widehat{P}$, and to each reflexive truth set $F$ in $V^{(\widehat{P})}$ the Grothendieck topology $\mathbf{C}_F$ on $P$ defined by $S \in \mathbf{C}_T(p) \leftrightarrow p \in [\![\hat{S} \in T]\!]$.

The topos $\mathscr{S}\!et^{(\bar{P})}$ of sets in $V^{(\bar{P})}$ is, as we have observed, equivalent to the topos of canonical sheaves on $\widehat{P}$, which is itself, as is well known, equivalent to the topos $\mathscr{S}\!et^{P^{op}}$ of presheaves on $P$. Moreover, Grothendieck topologies on $P$ are known (see [4]) to correspond bijectively to internal Lawvere-Tierney topologies—that is, internal nuclei—on the truth-value object $\Omega$ in $\mathscr{S}\!et^{P^{op}}$. How this fact related to the representation of Grothendieck topologies as reflexive truth sets in $V^{(\bar{P})}$? It turns out that in a general $V^{(H)}$ there is a natural bijection between truth sets/reflexive truth sets and weak nuclei/nuclei on $\Omega = \{u: u \subseteq \hat{1}\}$. The representation of Grothendieck topologies as truth sets in $V^{(H)}$, while equivalent to that through Lawvere topologies, seems especially perspicuous.

The *forcing* relation $\Vdash_P$ in $V^{(\bar{P})}$ between sentences and elements of $P$ is defined by

$$p \Vdash_P \sigma \leftrightarrow p \in [\![\sigma]\!]^{\bar{P}}.$$

This satisfies the usual rules governing Kripke semantics for predicate sentences, viz.,

- $p \Vdash_P \varphi \wedge \psi \leftrightarrow p \Vdash_P \varphi \ \& \ p \Vdash_P \psi$
- $p \Vdash_P \varphi \vee \psi \leftrightarrow p \Vdash_P \varphi \ \text{or} \ p \Vdash_P \psi$
- $p \Vdash_P \varphi \to \psi \leftrightarrow \forall q \leq p [ q \Vdash_P \varphi \to q \Vdash_P \psi]$
- $p \Vdash_P \neg \varphi \leftrightarrow \forall q \leq p \ q \not\Vdash_K \varphi$
- $p \Vdash_P \forall x \varphi \leftrightarrow p \Vdash_P \varphi(a)$ for every $a \in V^{(\bar{P})}$
- $p \Vdash_P \exists x \varphi \leftrightarrow p \Vdash_P \varphi(a)$ for some $a \in V^{(\bar{P})}$.

If **C** be a pretopology on $P$, the forcing relation $\Vdash_{\mathbf{C}}$ in the model $V^{(\bar{\mathbf{C}})}$ is similarly defined by

$$p \Vdash_{\mathbf{C}} \sigma \leftrightarrow p \in [\![\sigma]\!]_{\bar{\mathbf{C}}}.$$

As for Kripke models, this relation can be shown to satisfy the rules of Beth-Kripke-Joyal semantics, viz.,

- $p \Vdash_{\mathbf{c}} \varphi \wedge \psi \leftrightarrow p\Vdash_{\mathbf{c}} \varphi \ \& \ p \Vdash_{\mathbf{c}} \psi$
- $p\Vdash_{\mathbf{c}} \varphi \vee \psi \leftrightarrow \exists S \in \mathbf{C}(p) \ \forall s \in S \ [s\Vdash_{\mathbf{c}} \varphi \ \text{or} \ s\Vdash_{\mathbf{c}} \psi]$
- $p \Vdash_{\mathbf{c}} \varphi \Rightarrow \psi \leftrightarrow \forall q \leq p[\ q\Vdash_{\mathbf{c}} \varphi \ \rightarrow \ q\Vdash_{\mathbf{c}} \psi]$
- $p\Vdash_{\mathbf{c}} \neg\varphi \leftrightarrow \forall q \leq p \ [q\Vdash_{\mathbf{c}} \varphi \rightarrow \varnothing \in \mathbf{C}(p)]$
- $p\Vdash_{\mathbf{c}} \forall x \varphi \leftrightarrow p \Vdash_{\mathbf{c}} \varphi(a) \ \text{for every} \ a \in V^{(\overline{\mathbf{C}})}$
- $p\Vdash_{\mathbf{c}} \exists x \varphi \leftrightarrow \exists S \in \mathbf{C}(p) \ \forall s \in S \ s \Vdash_{\mathbf{c}} \varphi(a) \ \text{for some} \ a \in V^{(\overline{\mathbf{C}})}.$

## V. POTENTIAL APPLICATIONS OF COVER SCHEMES, KRIPKE MODELS, AND FRAME-VALUED SET THEORY IN SPACETIME PHYSICS

In spacetime physics any set $\mathcal{C}$ of events—a *causal set*—is taken to be partially ordered by the relation $\leq$ of *possible causation*: for $p, q \in \mathcal{C}$, $p \leq q$ means that $q$ is in $p$'s future light cone. In her groundbreaking paper [5] Fotini Markopoulou proposes that the causal structure of spacetime itself be represented by "sets evolving over $\mathcal{C}$" —that is, in essence, by the topos $\mathcal{S}\mathit{et}$ of presheaves on $\mathcal{C}^{\text{op}}$. To enable what she has done to be the more easily expressed within the framework presented here, we will reverse the causal ordering, that is, $\mathcal{C}$ will be replaced by $\mathcal{C}^{\text{op}}$, and the latter written as $P$—which will, moreover, be required to be no more than a *preordered* set. Specifically, then: $P$ is a set of events preordered by the relation $\leq$, where $p \leq q$ is intended to mean that $p$ is in $q$'s future light cone—that $q$ *could* be the cause of $p$. In requiring that $\leq$ be no more than a preordering—in dropping, that is, the antisymmetry of $\leq$—we are, in physical terms, allowing for the possibility that the universe is of Gödelian type, containing closed timelike lines.

Accordingly we fix a preordered set $(P, \leq)$, which we shall call the *universal causal set.* Markopoulou, in essence, suggests that viewing the universe "from the inside" amounts to placing oneself within the topos of presheaves $\mathit{Set}^{P^{op}}$. Since, as we have already observed, $\mathit{Set}^{P^{op}}$ is equivalent to the topos of sets in $V^{(\bar{P})}$, Markopoulou's proposal may be effectively realized by working within $V^{(\hat{P})}$. Let us do so, writing for simplicity $H$ for $\hat{P}$.

Define the set $K \in V^{(H)}$ by $\mathrm{dom}(K) = \{\hat{p} : p \in P\}$ and $K(\hat{p}) = p\!\downarrow$. Then, in $V^{(H)}$, $K$ is a subset of $\hat{P}$ and for $p \in P$, $[\![\hat{p} \in K]\!] = p\!\downarrow$. $K$ is the counterpart in $V^{(H)}$ of the evolving set *Past* Markopoulou defines by $Past(p) = p\!\downarrow$, with insertions as transition maps. ($\hat{P}$, incidentally, is the $V^{(H)}$-counterpart of the constant presheaf on $P$ with value $P$ —which Markopoulou calls *World*.) Accordingly the "causal past" of any "event" $p$ is represented by the truth value in $V^{(H)}$ of the statement $\hat{p} \in K$. The fact that, for any $p, q \in P$ we have

$$q \Vdash_P \hat{p} \in K \leftrightarrow q \leq p$$

may be construed as asserting that *the events in the causal future of an event p are precisely those forcing (the canonical representative of) p to be a member of K*. For this reason we shall call $K$ the *causal set in* $V^{(H)}$.

If we identify each $p \in P$ with $p\!\downarrow \in H$, $P$ may then be regarded as a subset of $H$ so that, in $V^{(H)}$, $\hat{P}$ is a subset of $\hat{H}$. It is not hard to show that, in $V^{(H)}$, $K$ generates the canonical prime filter $\Phi_H$ in $\hat{H}$. Using the V-genericity of $\Phi_H$, and the density of $P$ in $H$, one can show that $[\![\sigma]\!] = [\![\exists p \in K. p \leq \widehat{[\![\sigma]\!]}]\!]$, so that, with moderate abuse of notation,

$$V^{(H)} \vDash [\sigma \leftrightarrow \exists p \in K.\ p \Vdash \sigma].$$

That is, in $V^{(H)}$, a sentence holds precisely when it is forced to do so at some "causal past stage" in $K$. This establishes the centrality of $K$—and, correspondingly, that of the "evolving" set *Past*— in determining the truth of sentences "from the inside", that is, inside the universe $V^{(H)}$.

Markopoulou also considers the *complement* of *Past*—i.e., in the present setting, the $V^{(H)}$-set $\neg K$ for which $[\![\hat{p} \in \neg K]\!] = [\![p \notin K]\!] = \neg p{\downarrow}$. Markopoulou calls (*mutatis mutandis*) the events in $\neg p{\downarrow}$ those *beyond p's causal horizon*, in that no observer at $p$ can ever receive "information" from any event in $\neg p{\downarrow}$. Since clearly we have it follows that *the events beyond the causal horizon of an event p are precisely those forcing (the canonical representative of) p to be a member of $\neg K$*. In this sense $\neg K$ reflects, or "measures" the causal structure of $P$.

(*) $\qquad\qquad\qquad q \Vdash_P \hat{p} \in \neg K \;\leftrightarrow\; q \in \neg p{\downarrow},$

In this connection it is natural to call $\neg\neg p{\downarrow} = \{q : \forall r \leq q \exists s \leq r. s \leq p\}$ the *causal horizon* of $p$: it consists of those events $q$ for which an observer placed at $p$ could, in its future, receive information from any event in the future of an observer placed at $q$. Since

$$q \Vdash_P \hat{p} \in \neg\neg K \;\leftrightarrow\; q \in \neg\neg p{\downarrow},$$

it follows that *the events within the causal horizon of an event are precisely those forcing (the canonical representative of) p to be a member of $\neg\neg K$*.

It is easily shown that $\neg K$ is *empty* (i.e. $V^{(H)} \vDash \neg K = \varnothing$) if and only if $P$ is *directed downwards*, i.e., for any $p, q \in P$ there is $r \in P$ for which $r \leq p$ and $r \leq q$. This holds in the case, considered by Markopoulou, of *discrete Newtonian time evolution*—in the present setting, the case in which $P$ is the opposite $\mathbb{N}^{op}$ of the totally ordered set $\mathbb{N}$ of natural numbers. Here the corresponding complete Heyting algebra $H$ is the family of all downward-closed sets of natural numbers. In this case the $H$-valued set $K$ representing *Past is neither finite nor actually infinite in $V^{(H)}$*.

To see this, first note that, for any natural number $n$, we have, $[\![\neg(\hat{n} \in \neg K)]\!] = \mathbb{N}$. It follows that $V^{(H)} \vDash \neg\neg \forall n \in \hat{\mathbb{N}}. n \in K$. But, working in $V^{(H)}$, if $\forall n \in \widehat{\mathbb{N}}.\ n \in K$, then $K$ is not finite, so if $K$ is finite, then $\neg \forall n \in \widehat{\mathbb{N}}.\ n \in K$, and so $\neg\neg \forall n \in \widehat{\mathbb{N}}.\ n \in K$ implies the non-finiteness of $K$.

But, in $V^{(H)}$, $K$ is not actually infinite. For (again working in $V^{(H)}$), if $K$ were actually infinite (i.e., if there existed an injection of $\widehat{\mathbb{N}}$ into $K$), then the statement

$$\forall x \in K\ \exists y \in K.\ x > y$$

would also have to hold in $V^{(H)}$. But calculating that truth value gives:

$$\llbracket \forall x \in K \exists y \in K. x > y \rrbracket$$
$$= \bigcap_{m \in \mathbb{N}^{op}} [m \downarrow \Rightarrow \bigcup_{n \in \mathbb{N}^{op}} n \downarrow \cap \llbracket \hat{m} > \hat{n} \rrbracket]$$
$$= \bigcap_{m} [m \downarrow \Rightarrow \bigcup_{n < m} n \downarrow]$$
$$= \bigcap_{m} [m \downarrow \Rightarrow (m+1) \downarrow]$$
$$= \bigcap_{m} (m+1) \downarrow = \varnothing$$

So $\forall x \in K\ \exists y \in K.\ x > y$ is false in $V^{(H)}$ and therefore $K$ is not actually infinite. In sum, the causal set $K$ in is *potentially, but not actually infinite.*

In order to formulate an observable causal *quantum theory* Markopoulou considers the possibility of introducing a *causally evolving algebra of observables*. This amounts to specifying a presheaf of $C^*$-algebras on $P$, which, in the present framework, corresponds to specifying a set $\mathscr{A}$ in $V^{(H)}$ satisfying

$$V^{(H)} \vDash \mathscr{A} \text{ is a } C^*\text{-algebra.}$$

The "internal" $C^*$-algebra $\mathscr{A}$ is then subject to the intuitionistic internal logic of $V^{(H)}$: *any* theorem concerning $C^*$-algebras—provided only that it be constructively proved—automatically applies to $\mathscr{A}$. Reasoning with $\mathscr{A}$ is more direct and simpler than reasoning with $A$.

This same procedure of "internalization" can be performed with any causally evolving object: each such object of type $\mathscr{T}$ corresponds to a set $S$ in $V^{(H)}$ satisfying

$$V^{(H)} \models \ S \text{ is of type } \mathcal{T}.$$

Internalization may also be applied in the case of the presheaves *Antichains* and *Graphs* considered by Markopoulou. Here, for each event *p*, *Antichains*(*p*) consists of all sets of causally unrelated events in *Past*(*p*), while *Graphs*(*p*) is the set of all graphs supported by elements of *Antichains*(*p*). In the present framework *Antichains* is represented by the $V^{(H)}$–set *Anti* = { $X \subseteq \hat{P}$: *X is an antichain*} and *Graphs* by the $V^{(H)}$–set *Grph* = {*G*: ∃*X* ∈ *A* .*G is a graph supported by A*}. Again, both *Anti* and *Grph* can be readily handled using the internal intuitionistic logic of $V^{(H)}$.

What are the possible uses of *cover schemes* here? As we have seen, cover schemes may be used to force certain conditions to prevail in the associated models. Let us consider, for example, the cover scheme **Den** in *P*. We know that the associated frame $\widehat{\textbf{Den}}$ is a Boolean algebra— let us denote it by *B*. The corresponding causal set $K_B$ in $V^{(B)}$ then has the property

$$[\![ \hat{p} \in K_B ]\!] = \neg\neg p\!\downarrow;$$

so that,

$$q \Vdash_B \hat{p} \in K_B \leftrightarrow q \in \neg\neg p\!\downarrow$$

$$\leftrightarrow q \text{ is in } p\text{'s causal horizon.}$$

Comparing this with (*) above, we see that moving to the universe $V^{(B)}$—"Booleanizing" it, so to speak—*amounts to replacing causal futures by causal horizons.* When *P* is linearly ordered, as for example in the case of Newtonian time, the causal horizon of any event coincides with the whole of *P*, *B* is the two-element Boolean algebra **2,** so that $V^{(B)}$ is just the universe *V* of "static" sets. In this case, then, the effect of "Booleanization" is to *render the universe timeless.*

The universes associated with the cover schemes $\mathbf{C}^A$ and $\mathbf{C}_A$ seem also to have a rather natural physical meaning. Consider, for instance the case in which $A$ is the sieve $p{\downarrow}$—the causal future of $p$. In the associated universe $V^{(\widehat{\mathbf{C}^A})}$ the corresponding causal set $K^A$ satisfies

$$[\![\hat{q} \in K^A]\!] = \text{least } \mathbf{C}^A\text{-closed sieve containing } q$$

so that, in particular

$$[\![\hat{p} \in K^A]\!] = \text{least } \mathbf{C}^A\text{-closed sieve containing } p$$
$$= P.$$

This means that, for every event $q$,

$$q \Vdash_{\widehat{\mathbf{C}^A}} \hat{p} \in K^A.$$

Comparing this with (*), we see that in $V^{(\widehat{\mathbf{C}^A})}$ that every event has been "forced" into $p$'s causal future: in short, that $p$ now marks the "beginning" of the universe as viewed from inside $V^{(\widehat{\mathbf{C}^A})}$.

Similarly, we find that the causal set $K_A$ in the universe $V^{(\widehat{\mathbf{C}_A})}$ has the property

$$q \leq p \rightarrow \forall r[r \Vdash_{\widehat{\mathbf{C}_A}} \hat{q} \in \neg K_A];$$

that is, $p$, together with all events in its causal future, are, in $V^{(\widehat{\mathbf{C}^A})}$, beyond the causal horizon of any event. In effect, $p$ has become a *black hole*.

As a final possibility consider the universe $V^{(\widetilde{P})}$ associated with the free lower semilattice $\widetilde{P}$ generated by $P$. In this case the elements of $\widetilde{P}$ are finite sets of events, preordered by the relation $\sqsubseteq$: for $F, G \in \widetilde{P}$, $F \sqsubseteq G$ iff every event in $G$ is in the causal past of an event in $F$. The empty set of events is the top element of $\widetilde{P}$. The causal set $\widetilde{K}$ in $V^{(\widetilde{P})}$ has the property that its complement $\neg\widetilde{K}$ is empty and $\hat{\varnothing}$ is an initial event in the sense that $F \Vdash_{\tilde{P}} \widehat{\varnothing} \in \widetilde{K}$ for every "event" $F$. In this case passage to the new universe $V^{(\tilde{\bar{P}})}$ preserves the original causal relations in the sense that

$$\{q\} \Vdash_{\tilde{P}} \widehat{\{p\}} \in \widetilde{K} \leftrightarrow q \Vdash_{P} \widehat{p} \in K,$$

and so the initial event $\widehat{\varnothing}$ has been "freely adjoined" to $V^{(\tilde{P})}$.

## REFERENCES

[1] Bell, J.L. *Boolean-Valued Models and Independence Proofs in Set Theory.* Oxford University Press, 1977.

[2] Grayson, R. J. *Heyting-valued models for intuitionistic set theory.* In: *Applications of Sheaves.* Springer Lecture Notes in Mathematics **753**, 1979.

[3] Johnstone, P.T. *Stone Spaces.* Cambridge University Press, 1982.

[4] Mac Lane, S. and Moerdijk, I. *Sheaves in Geometry and Logic.* Springer-Verlag, 1992.

[5] Markopoulou, F. *The internal description of a causal set: What the universe looks like from the inside.* arXiv:gr-qc/98ii053 v2  18 Nov 1999.